\documentclass[multphys,vecphys]{svmult}

\usepackage{makeidx}         
\usepackage{graphicx}        
\usepackage{multicol}        
\usepackage[bottom]{footmisc}

\newcommand{\gsim}{\raisebox{-0.13cm}{~\shortstack{$>$ \\[-0.07cm] $\sim$}}~}

\makeindex             

\begin{document}

\title*{The role of {\em HST} in the study of near- and mid-infrared-selected galaxies}


\author{Karina I. Caputi \inst{1}}


\institute{Institute for Astronomy, Swiss Federal Institute of Technology (ETH H\"onggerberg), 
Wolfgang-Pauli-Strasse 16, CH-8093, Z\"urich, Switzerland.  
\texttt{caputi@phys.ethz.ch}
}
%
%
\maketitle

\begin{abstract}
Because of their unique quality, {\em Hubble Space Telescope (HST)} data have played an important complementary role in studies of infrared (IR) galaxies conducted with major facilities, as {\em VLT} or {\em Spitzer}, and will be as well very valuable for future telescopes as {\em Herschel} and {\em ALMA}.  I review here some of the most recent works led by European astronomers on IR galaxies, and discuss the role that {\em HST} has had in the study of different IR galaxy populations. I particularly focus the analysis on the GOODS fields, where the multiwavelength data and unique {\em HST}  coverage have enabled to jointly put constraints on the evolution of star formation activity and stellar-mass growth with cosmic time.
\end{abstract}

\section{Introduction}
\label{sec:1}


Studies of galaxy populations much benefit from the coordinated efforts of multiwavelength observations, which allow for a deeper insight into different galaxy properties and evolution. One of the best examples of these coordinated efforts is the {\em Great Observatories Origins Deep Survey (GOODS)} (P.I.: M. Giavalisco), with  320 arcmin$^2$ of the sky fully covered with deep observations from X-rays through radio wavelengths. Both North and South GOODS fields have been designed to be a unique case of {\em HST} coverage: they have deep and homogeneous maps in four broad-band {\em Advanced Camera for Surveys (ACS)} filters $B, V, I_{775}$ and $z_{850}$. The large amount of science produced over the last years using GOODS data demonstrates the importance of {\em HST} Treasury and other major-telescope Legacy programs.

Near and mid-IR selected galaxies constitute quite unbiased tracers of two different astrophysical properties: stellar mass and star-formation/AGN activity, respectively.   Although, in general, these galaxies populations are studied separately, a joint analysis of results allows to better understand the interplay between star formation, AGN activity and stellar-mass growth at different cosmic times. In the study of near- and mid-IR-selected galaxies, deep {\em HST} data have had a major role in the following aspects: 1) the determination of  optical spectral energy distributions (SEDs), 2) the computation of precise ($\sigma \approx 3-5\%$) photometric redshifts, as a complement to existing spectroscopic redshifts and 3) galaxy morphology. The two former aspects have been of key importance to conduct complete studies of galaxy luminosity evolution and stellar mass assembly.

\section{Near-IR-selected galaxies}
\label{sec:2}

An important landmark in the study of near-IR-selected galaxies has been the K20 survey \cite{cim02}, which included part of the GOODS-South among their fields. The first science results obtained by the K20 survey have soon been extended after the progressive public releases of the deeper {\em ISAAC-VLT} $J_s, H$ and $K_s$-band data in the GOODS-South (P.Is.: C. Cesarsky, E. Giallongo). Finally, after the launch of {\em Spitzer} in 2003 (P.I.: M. Werner), both GOODS fields have been observed with {\em IRAC} at 3.6, 4.5, 5.8 and 8.0 $\rm \mu m$ (P.I.: M. Dickinson).

At the typical depths of the GOODS data (21-22 Vega mag), $K_s$-selected galaxies mostly span the redshift range $z=0-1.5$, but  $\sim 20\%$ of these galaxies are found at higher $z=1.5-5.0$ redshifts \cite{cap05}\cite{cap06k}\cite{gra06}. The latter include Extremely Red Galaxies (ERGs),  which have been the subject of many studies looking for the progenitors of present-day massive ellipticals \cite{cap05}\cite{smail02}\cite{roch03}\cite{cap04}\cite{roch06}.

\begin{figure}
\centering
\includegraphics[width=9cm,angle=270]{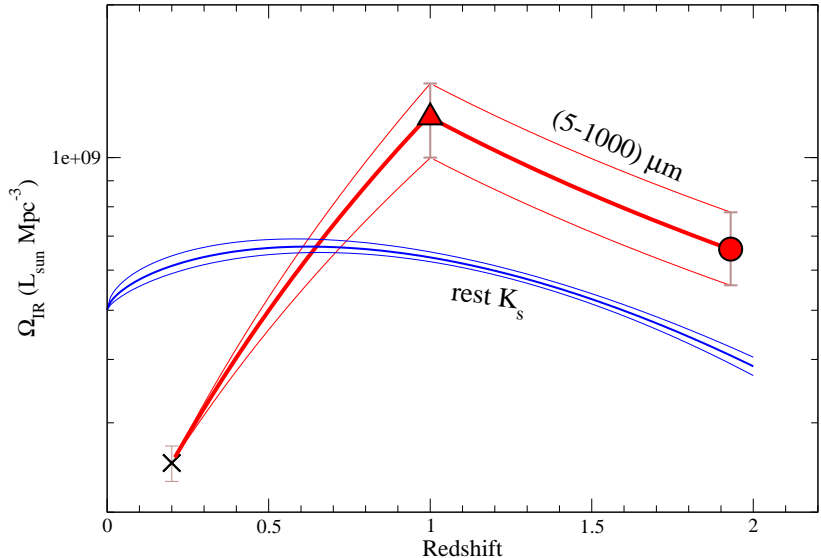}
\caption{The evolution of the IR luminosity densities associated with the stars already locked in galaxies (rest-$K_s$) and with on-going star formation ($5-1000 \, \rm \mu m$). The cosmology adopted has $\rm H_0=70 \, km \,s^{-1}\, Mpc^{-1}$, $\rm \Omega_{\rm M}=0.3$ and $\rm \Omega_{\rm \Lambda}=0.7$.}
\label{fig1}       
\end{figure}

The evolution of the rest-frame $K_s$-band luminosity function (LF) has been studied up to $z \approx 1.5$ with the K20 survey \cite{pozz03}, and then extended up to redshift $z \approx 2.5$ using GOODS and other datasets \cite{cap05} \cite{cap06k} \cite{sar06}\cite{arn07}. All these works agree in a  mild but still significant increase of the bright end of this LF from $z\approx0$ to $z\approx2.0-2.5$. At the same time, the overall number density of galaxies decreases,  making the near-IR luminosity density to be nearly constant with redshift (cf. Figure \ref{fig1}).

Multiwavelength SED modelling has also allowed to obtain stellar-mass estimates for near-IR-selected galaxies. The redshift evolution of galaxy number densities is differential with stellar mass: while the density of moderately massive galaxies ($M\sim 1 \times10^{11} \, \rm M_\odot$) continuosly increases from redshift $z\sim4$ down to $z\sim0.5$, virtually all of the most massive ($M>2-3 \times 10^{11} \, \rm M_\odot$) galaxies appear to be in place by $z\sim1.5-2.0$ \cite{cap06k}\cite{sar05}\cite{tho05}\cite{fran06}\cite{ren06}\cite{pozz07}. The formation epoch of the most massive galaxies seems to be constrained to the redshift range $z\sim 2-4$, as no evidence of such galaxies has been found at $z>4$ in the GOODS-South field, indicating that these objects must be very rare (if any exists) at very high redshifts \cite{dun07}.

Moreover, during the last years, a general consensus has  been achieved on the redshift evolution of the total stellar mass density:  half of the stellar mass has been assembled in galaxies before the Universe was 30-40\% of its present age \cite{cap05}\cite{cap06k}\cite{fon04}.

%
%
%

\section{Mid-IR-selected galaxies and their link to near-IR galaxies}
\label{sec:3}

The study of mid-IR selected galaxies has been a subject of major interest in European astronomy since the launch of the {\em Infrared Space Observatory (ISO)} e.g. \cite{aus99}\cite{gen00}\cite{fra01}.   The amount of energy contained in the extragalactic IR background ($\lambda \approx 5-1000 \, \rm \mu m$) is comparable to that in the optical background \cite{dole06}, which emphasizes the importance of understanding the role of mid- and far-IR galaxy populations  in the general context of galaxy evolution. 

After {\em ISO} set the first constraints on the mid-IR Universe up to redshift $z\sim1$,  {\em Spitzer} has allowed us  to prove that mid-IR galaxies have in fact a significant contribution up to higher $z\sim2-3$ redshifts, e.g.\cite{cap06b}. A sort of `downsizing' effect is observed in the typical stellar masses of the galaxies hosting the bulk of IR activity: at $z\sim2$, most of the IR activity is found in $\gsim 10^{11} \, \rm M_\odot$ galaxies, while at $z=0.5-1.0$, an important fraction of the IR activity is concentrated in intermediate-mass galaxies, with  $10^{10} < \rm M < 10^{11} \, M_\odot$ \cite{cap06b}\cite{ham05}.

A result that might have come as a surprise is the large fraction of massive galaxies that are experiencing an ultra-luminous IR phase at high redshifts \cite{daddi05}\cite{cap06c}. Very recent results suggest that the AGN contribution to this IR phase at $z\sim2$ might be much more important than previously thought \cite{daddi07}.

The mid-IR LF strongly evolves from $z=0$ to 1 \cite{lefl05}\cite{cap07}. But, after removing to our best current knowledge the AGN contribution at high redshifts, the mid-IR LF associated with star formation shows only a modest evolution from $z=1$ to $z=2$ \cite{cap07}. This is reflected in the observed evolution of the bolometric IR  ($5-1000 \, \rm \mu m$) luminosity density, as shown in Figure \ref{fig1}. From this figure, it is clear that the amount of energy associated with on-going star formation is significantly larger than the light produced by the stars already locked in galaxies at $z\gsim 0.6-0.7$, while the balance is reversed at lower redshifts.  The  $(5-1000) \, \rm \mu m$ luminosity related to star formation can be interpreted as the gradient of the evolution of the assembled stellar mass. By $z\sim0.7$, 80\% of the present-day stellar mass is already in place \cite{cap06k}, so there is only a minor amount of star formation still needed afterwards to finish assembling our local Universe.



\printindex

\begin{thebibliography}{99.}
\bibitem{cim02} Cimatti, A., Mignoli, M., Daddi, E., et al., 2002, A\&A, 392, 395 
\bibitem{cap05} Caputi K.I., Dunlop, J.S., McLure, R.J., Roche, N.D., 2005, MNRAS, 361, 607
\bibitem{cap06k} Caputi K.I., McLure, R.J., Dunlop J.S., et al., 2006, MNRAS, 366, 609
\bibitem{gra06} Grazian, A., Fontana, A., de Santis, C., et al., 2006, A\&A, 449, 951
\bibitem{smail02} Smail I., Owen, F.N., Morrison, G.E., et al., 2002, ApJ, 581, 844 
\bibitem{roch03} Roche, N.D., Dunlop, J.S., Almaini, O., 2003, MNRAS, 346, 803
\bibitem{cap04} Caputi K.I., Dunlop, J.S., McLure, R.J., Roche, N.D., 2004, MNRAS, 353, 30
\bibitem{roch06} Roche, N.D., Dunlop J.S., Caputi K.I., et al., 2006, MNRAS, 370, 74
\bibitem{pozz03} Pozzetti, L., Cimatti, A., Zamorani, G., et al., 2003, A\&A, 402, 837 
\bibitem{sar06}  Saracco, P., Fiano, A., Chincarini, G., et al., 2006, MNRAS, 367, 349
\bibitem{arn07} Arnouts, S., Walcher, C.J., Le F\`evre, O., et al., 2007, A\&A, submitted (arXiv:0705.2438) 
\bibitem{sar05} Saracco, P., Longhetti, M., Severgnini, P., et al., 2005, MNRAS, 357, L40
\bibitem{tho05} Thomas, D., Maraston, C., Bender, R., et al., 2005, ApJ, 621, 673
\bibitem{fran06} Franceschini, A., Rodighiero, G., Cassata, P., et al., 2006, A\&A, 453, 397
\bibitem{ren06} Renzini, A., 2006, ARA\&A, 44, 141
\bibitem{pozz07} Pozzetti, L., Bolzonella, M., Lamareille, F., et al., 2007, A\&A, submitted (arXiv:0704.1600)
\bibitem{dun07} Dunlop, J.S., McLure, R.J., Cirasuolo, M., 2007, MNRAS, 376, 1054 
\bibitem{fon04} Fontana, A., Pozzetti, L., Donnarumma, I., et al., 2004,  A\&A, 424, 23
\bibitem{aus99} Aussel, H., Cesarsky, C.J., Elbaz, D., et al., 1999, A\&A., 342, 313
\bibitem{gen00} Genzel, R. \& Cesarsky, C.J., 2000, ARA\&A, 38, 761
\bibitem{fra01} Franceschini, A., Aussel, H., Cesarsky, C.J., et al., 2001,  A\&A., 378, 1
\bibitem{dole06} Dole, H., Lagache, G., Puget, J.-L., et al., 2006, A\&A, 451, 417
\bibitem{cap06b} Caputi K.I., Dole, H., Lagache, G., et al., 2006, ApJ, 637, 727
\bibitem{ham05} Hammer F., Flores, H., Elbaz, D., et al., 2005, A\&A, 430, 115
\bibitem{daddi05} Daddi, E., Dickinson, M., Chary, R., et al., 2005, ApJ, 631, L13
\bibitem{cap06c} Caputi, K.I., Dole, H., Lagache, G., et al., 2006, A\&A, 454, 143
\bibitem{daddi07} Daddi, E., Alexander, D. M. , Dickinson, M., et al., 2007, ApJ, submitted (arXiv:0705.2832)
\bibitem{lefl05} Le Floc'h, E., Papovich, C., Dole, H., et al., 2005, ApJ, 632, 169
\bibitem{cap07} Caputi, K.I., Lagache, G., Yan, L., et al., 2007, ApJ, 660, 97
\end{thebibliography}
\end{document}